\documentstyle[prd,aps,eqsecnum,twocolumn,floats]{revtex}

\def\beq{\begin{equation}}
\def\eeq{\end{equation}}
\def\beqa{\begin{eqnarray}}
\def\eeqa{\end{eqnarray}}
\def\bfig{\begin{figure}}
\def\efig{\end{figure}}

\input{psfig}
\begin{document}
\draft
\fnsymbol{footnote}

\wideabs{

\title{Generalized {\it r-}Modes of the Maclaurin Spheroids}

\author{Lee Lindblom}
\address{Theoretical Astrophysics 130-33,
         California Institute of Technology,
         Pasadena, CA 91125}

\author{James R. Ipser}
\address{Department of Physics,
         University of Florida,
         Gainesville, FL 32611}

\date{\today}

\maketitle

\begin{abstract}

Analytical solutions are presented for a class of generalized
$r-$modes of rigidly rotating uniform density stars---the Maclaurin
spheroids---with arbitrary values of the angular velocity.  Our
analysis is based on the work of Bryan~\cite{bryan}; however, we
derive the solutions using slightly different coordinates that give
purely real representations of the $r-$modes.  The class of
generalized $r-$modes is much larger than the previously studied
`classical' $r-$modes.  In particular, for each $l$ and $m$ we find
$l-m$ (or $l-1$ for the $m=0$ case) distinct $r-$modes.  Many of these
previously unstudied $r-$modes (about $30\%$ of those examined) are
subject to a secular instability driven by gravitational radiation.
The eigenfunctions of the `classical' $r-$modes, the $l=m+1$ case
here, are found to have particularly simple analytical
representations.  These $r-$modes provide an interesting mathematical
example of solutions to a hyperbolic eigenvalue problem.

\pacs{PACS Numbers: 97.10.Sj, 97.10.Kc, 04.30.Db}
\end{abstract}

}

\narrowtext

\section{Introduction}
\label{section0}

During the past year the $r-$modes of rotating neutron stars have been
found to play in interesting and important role in relativistic
astrophysics.  Andersson~\cite{andersson} and Friedman and
Morsink~\cite{fried-morsink} showed that these modes would be driven
unstable by gravitational radiation reaction in the absence of internal
fluid dissipation.  Lindblom, Owen, and Morsink~\cite{LOM} have
subsequently shown that this instability will in fact play an important
role in the evolution of hot young neutron stars.  The gravitational
radiation reaction force in these modes was shown to be sufficiently
strong to overcome the internal fluid dissipation present in neutron
stars hotter than about $10^9\,$K.  Hot young rapidly rotating neutron
stars are expected therefore to radiate away most (i.e.  up to about
$90\%$) of their angular momentum via gravitational radiation in a
period of about one year.  Owen, et al.~\cite{owen} have shown that the
gravitational radiation emitted during this spin-down process is expected
to be one of the more promising potential sources for the ground based
laser interferometer gravitational wave detectors (e.g.  LIGO, VIRGO,
etc.) now under construction. 

To date the various analyses of the $r-$modes and their instability to
gravitational radiation reaction have all been based on small angular
velocity approximations.  This instability is of primary importance in
astrophysics for rapidly rotating stars.  The purpose of this paper is
to provide the first look at the properties of these important modes
in stars of large angular velocity.  We do this by solving the stellar
pulsation equations for the $r-$modes of the rapidly rotating uniform
density stellar models which are known as the Maclaurin spheroids.
The pulsations of these models were studied over a century ago by
Bryan~\cite{bryan}, who showed how analytical expressions for all of
the modes of these stars could be found.  We follow the general
strategy developed by Bryan to derive analytical expressions for the
$r-$modes of these stars.  We use somewhat different coordinates than
Bryan, however, in order to obtain real representations of the
$r-$modes (of primary interest to us here) using purely real
coordinates.

We generalize the traditional definition of $r-$mode to include any
mode whose frequency vanishes linearly with the angular velocity of
the star.  Such modes have as their principal restoring force the
Coriolos force, and hence it is appropriate to call them rotation
modes or generalized $r-$modes~\cite{def}.  We find a very large
number of generalized $r-$modes in the Maclaurin
spheroids~\cite{g-modes}.  In particular for each pair of integers $l$
and $m$ (with $l\geq m\geq 0$) we find $l-m$ (or $l-1$ for the $m=0$
case) distinct $r-$modes.  The `classical' $r-$modes as studied for
example by Papaloizou and Pringle~\cite{p&p} correspond here to the
case $l=m+1$~\cite{l=m+1}.  We find that many of the previously
unstudied $r-$modes (about $30\%$ of those examined) are also subject
to the gravitational radiation driven instability in stars without
internal fluid dissipation.  These new $r-$modes couple to higher
order gravitational multipoles and consequently are expected to be of
less astrophysical importance than the $l=m+1=3$ mode that is of
primary importance in the instabilty discussed by Lindblom, Owen, and
Morsink~\cite{LOM}.

In Section~\ref{sectionII} we review a few important facts about the
equilibrium structures of the Maclaurin spheroids.  In
Sec.~\ref{sectionI} we present the equations for the modes of rapidly
rotating stars using the two-potential formalism~\cite{ipser-lind}.
This formalism determines all of the properties of the modes of
rotating stars from a pair of scalar potentials: a hydrodynamic
potential $\delta U$ and the gravitational potential $\delta \Phi$.
The equations satisfied by these potentials are (coupled) second-order
partial differential equations with suitable boundary conditions at
the surface of the star and at infinity.  These equations become
extremely simple in the case of uniformly rotating uniform-density
stars.  In Sec.~\ref{sectionIII} we introduce coordinates which allow
the equations for the two potentials to be solved analytically.  And
following in the footsteps of Bryan~\cite{bryan}, we present the
general solutions to these equations for the generalized $r-$modes of
the Maclaurin spheroids.  In Sec.~\ref{sectionIV} we give expressions
for the various boundary conditions in the coordinates adapted to this
problem.  In Sec.~\ref{sectionV} we deduce the eigenvalue equation
that determines when the boundary conditions may be satisfied.  In
Sec.~\ref{sectionV} we also present explicit solutions to this
eigenvalue equation for a large number of generalized $r-$modes.  In
the limit of small angular velocity we tabulate a complete set of
solutions for all generalized $r-$modes with $l\leq 6$.  We also
present graphically the angular velocity dependence of each of these
$r-$modes which is unstable to the gravitational radiation
instability.  In Sec.~\ref{sectionVI} we analyze the analytical
expressions for the eigenfunctions of these modes.  We show that in
the case of the `classical' $r-$modes, the $l=m+1$ case here, the
eigenfunctions $\delta U$ and $\delta \Phi$ have particularly simple
forms.  In particular, each of these eigenfunctions is simply
$r^{m+1}Y_{m+1\,m}(\theta,\varphi)$ (where $r$, $\theta$ and $\varphi$
are the standard spherical coordinates) multiplied by some angular
velocity dependent normalization.  We also show that these `classical'
$r-$mode eigenfunctions have the unexpected property that $\Delta p$,
the Lagrangian pressure perturbation, vanishes identically throughout
the star.  In Sec.~\ref{sectionVII} we discuss some of the interesting
implications of this analysis.  And finally, in the Appendix we
explore in some detail the properties of the rather unusual
bi-spheroidal coordinate system needed in Sec.~\ref{sectionIII} to
solve the pulsation equations for the hydrodynamic potential $\delta
U$.

\section{The Maclaurin Spheroids}
\label{sectionII}

The uniformly rotating uniform-density equilibrium stellar models are
called Maclaurin spheroids.  The structures of these stars
are determined by solving the time independent Euler equation

\beq
0=\nabla_a\left[\case{1}{2}(x^2+y^2)\Omega^2+{p\over\rho}  -
\Phi\right].\label{2.0}
\eeq

\noindent In this equation $p$ is the pressure, $\rho$ is the density,
$\Omega$ is the angular velocity, and $\Phi$ is the gravitational
potential of the equilibrium star.  Using the expression for the
gravitational potential of a uniform-density spheroid~\cite{chandra}, it
is straightforward to show that the solution to Eq.~(\ref{2.0}) for
the pressure is

\beqa
p=2\pi G \rho^2 \zeta_o^2(1+\zeta_o^2)&&(1-\zeta_o\cot^{-1}\zeta_o)
\nonumber\\
&&\times\left(a^2-{x^2+y^2\over 1+\zeta_o^2}-{z^2\over \zeta_o^2}\right),
\label{2.2}
\eeqa

\noindent where $a$ is the focal radius of the spheroid, $G$ is Newton's
constant, and $\zeta_o$ is related to the eccentricity $e$ of the
spheroid by $e^2=1/(1+\zeta_o^2)$.  Similarly, it follows that
the angular velocity of the star is related to the shape of
the spheroid by

\beq \Omega^2 = 2\pi G \rho \,
\zeta_o \left[ (1+3\zeta_o^2)\cot^{-1}\zeta_o -3\zeta_o\right].
\label{2.1} \eeq

\noindent    We note that small
angular velocities, $\Omega$, correspond to small eccentricities $e$
and large $\zeta_o$. 

The surfaces of these stellar models are the surfaces on which the
pressure vanishes:

\beq
{x^2+y^2\over \zeta_o^2+1}+{z^2\over \zeta_o^2}=a^2.\label{2.3}
\eeq

\noindent This surface is an oblate spheroid.  Let us denote the
equatorial and polar radii of this spheroid as $R_e$ and $R_p$
respectively.  We see from Eq.~(\ref{2.3}) that

\beq
R_e^2 = a^2(\zeta_o^2+1),\label{2.3.1}
\eeq

\beq
R_p^2 = a^2 \zeta_o^2.\label{2.3.2}
\eeq

\noindent If we consider a sequence of uniformly rotating spheroids
having the same total mass, then the volume of each of the spheroids in
the sequence is the same (since the density is constant).  Let $R$
denote the average radius of the spheroid: $R^3=R_e^2R_p$.  It follows
that $R$ is constant along this sequence since the volume of a spheroid
is $V=4\pi R^3/3$.  Thus the angular velocity dependence of the focal
radius $a$ is determined by

\beq
a^3={R^3\over \zeta_o(\zeta_o^2+1)},\label{2.3.3}
\eeq

\noindent since $\zeta_o$ is related to the angular
velocity of the spheroid $\Omega$ by Eq.~(\ref{2.1}).  This
expression determines then the angular velocity dependencies
of the equatorial and polar radii of the spheroid:

\beq R_e = R \left({\zeta_o^2+1\over \zeta_o^2}\right)^{1/6},
\label{2.3.4} \eeq

\beq
R_p=R \left({\zeta_o^2\over \zeta_o^2+1}\right)^{1/3}.
\label{2.3.5}
\eeq

In the work that follows we will need the quantity $n^a\nabla_ap$, where
$n^a$ is the outward directed unit normal to the surface of the star, in
order to evaluate certain boundary conditions associated with the
stellar pulsations.  Since $\nabla_ap$ is also normal to the surface of
the star, we may use the expression $(n^a\nabla_ap)^2 =
\nabla^ap\nabla_ap$ and Eq.~(\ref{2.2}) to obtain

\beqa
n^a\nabla_ap=-4\pi G\rho^2\zeta_o^2(1&&+\zeta_o^2)
(1-\zeta_o\cot^{-1}\zeta_o)\nonumber\\
&&\times\left[{x^2+y^2\over (1+\zeta_o^2)^2}+{z^2\over \zeta_o^4}
\right]^{1/2},
\label{2.4}
\eeqa

\noindent where $(x,y,z)$ are to be confined to the surface defined
by Eq.~(\ref{2.3}).

\section{The Pulsation Equations} \label{sectionI}

The modes of any uniformly rotating barotropic stellar model are
determined completely in terms of two scalar potentials $\delta U$ and
$\delta\Phi$~\cite{ipser-lind}.  The potential $\delta \Phi$ is the
Newtonian gravitational potential, while $\delta U$ is a potential that
primarily describes the hydrodynamic perturbations of the star:

\beq
\delta U = {\delta p\over \rho}-\delta\Phi,\label{1.1}
\eeq

\noindent where $\delta p$ is the Eulerian pressure perturbation, and
$\rho$ is the unperturbed density of the equilibrium stellar model.  We
assume here that the time dependence of the mode is $e^{i\omega t}$ and
that its azimuthal angular dependence is $e^{im\varphi}$, where $\omega$
is the frequency of the mode and $m$ is an integer.  The velocity
perturbation $\delta v^a$ is determined by solving Euler's equation. 
This reduces in this case to an algebraic relationship between $\delta
v^a$ and the potential $\delta U$:

\beq
\delta v^a = iQ^{ab}\nabla_b\delta U.\label{1.2}
\eeq

\noindent The tensor $Q^{ab}$ depends on the frequency of the mode, and
the angular velocity of the equilibrium star $\Omega$:

\beqa
Q^{ab}=&&{1\over (\omega+m\Omega)^2-4\Omega^2}\nonumber\\
&&\times\Biggl[(\omega+m\Omega)\delta^{ab}-
       {4\Omega^2\over \omega+m\Omega}z^az^b - 2i\nabla^av^b\Biggr].
\label{1.3}
\eeqa

\noindent In Equation~(\ref{1.3}) the unit vector $z^a$ points along the
rotation axis of the equilibrium star, $\delta^{ab}$ is the Euclidean
metric tensor (the identity matrix in Cartesian coordinates), and $v^a$
is the velocity of the equilibrium stellar model.  The potentials
$\delta U$ and $\delta \Phi$ are determined then by solving the
perturbed mass conservation and gravitational potential equations.  In
this case, these reduce to the following system of partial differential
equations~\cite{ipser-lind}

\beq
\nabla_a(\rho Q^{ab}\nabla_b\delta U)=-(\omega+m\Omega)\rho
{d\rho\over dp}(\delta U+\delta\Phi),\label{1.4}
\eeq

\beq
\nabla^a\nabla_a\delta\Phi = -4\pi G\rho{d\rho\over dp}
(\delta U +\delta\Phi).\label{1.5}
\eeq

\noindent In addition these potentials are subject to the appropriate
boundary conditions at the surface of the star for $\delta U$ and at
infinity for $\delta \Phi$.  

The stellar pulsation Eqs.~(\ref{1.4}) and (\ref{1.5}) simplify
considerably for the case of uniformly rotating uniform-density stellar
models.  In this case $d\rho/dp=\rho\,\delta(p)$ [where $\delta(p)$ is
the Dirac delta function].  Thus the right sides of Eqs.~(\ref{1.4}) and
(\ref{1.5}) vanish except on the surface of the star.  Further, the
density $\rho$ that appears on the left side of Eq.~(\ref{1.4}) may be
factored out.  The resulting equations then in the uniform-density case
are simply

\beq
\kappa^2\nabla^a\nabla_a\delta U - 4z^az^b\nabla_a\nabla_b\delta U
=0,\label{1.6}
\eeq
\beq
\nabla^a\nabla_a\delta\Phi = -4\pi G \rho^2 \delta(p)
\,(\delta U + \delta\Phi)\,
,\label{1.7}
\eeq

\noindent where $\kappa$ is related to the frequency of the mode by

\beq
\kappa\Omega = \omega+m\Omega.\label{1.8}
\eeq

\noindent These equations are equivalent to those used by Bryan~\cite{bryan}
in his analysis of the oscillations of the Maclaurin spheroids.

Next we wish to consider the boundary conditions to which the functions
$\delta U$ and $\delta \Phi$ are subject.  Let $\Sigma$ denote a
function which vanishes on the surface of the star, and which has been
normalized so that its gradient, $n_a=\nabla_a\Sigma$, is the outward
directed unit normal vector there, $n^an_a=1$.  First, the function
$\delta U$ must be constrained at the surface of the star, $\Sigma=0$,
in such a way that the Lagrangian perturbation in the pressure vanishes
there, $\Delta p =0$.  This condition can be written in terms of the
variables used here by noting that

\beq
\Delta p = \delta p +\left({\delta v^a\over i\kappa \Omega}\right)\nabla_ap. 
\label{1.9}
\eeq

\noindent Then using Eqs.~(\ref{1.1}) and (\ref{1.2}) the boundary
condition can be written in terms of $\delta U$ and $\delta \Phi$ as

\beq
0=\biggl[\rho\,\kappa\Omega(\delta U +\delta\Phi)
+ Q^{ab}\nabla_a p \nabla_b\delta U
\biggr]_{\Sigma\uparrow 0}.\label{1.10}
\eeq

\noindent The perturbed gravitational potential $\delta\Phi$ must
vanish at infinity, $\lim_{\,r\rightarrow \infty}\delta\Phi = 0$.  In
addition $\delta\Phi$ must as a consequence of Eq.~(\ref{1.7}) have a
finite discontinuity in its first derivative at the surface of the
star.  In particular the derivatives must satisfy

\beq 
\biggl[n^a\nabla_a\delta\Phi\biggr]_{\Sigma\downarrow 0}
=\biggl[n^a\nabla_a\delta\Phi+ {4\pi G \rho^2(\delta U +
\delta\Phi) \over n^a\nabla_ap}\biggr]_{\Sigma\uparrow 0}.
\label{1.11}
\eeq

\noindent The problem of finding the modes of uniform-density stars is
reduced therefore to finding the solutions to Eqs.~(\ref{1.6}) and (\ref{1.7})
subject to the boundary conditions in Eqs.~(\ref{1.10}) and (\ref{1.11}).

\section{Solving For The Potentials}
\label{sectionIII}

In this section we find the general solutions for the potentials
$\delta U$ and $\delta\Phi$ that satisfy Eqs.~(\ref{1.6}) and
(\ref{1.7}).  This analysis basically follows that of
Bryan~\cite{bryan} except for some changes to modernize notation, and
a change of coordinates to express in a purely real manner the
solutions of interest to us here.  Our primary concern here is to find
expressions for the generalized $r-$modes of the Maclaurin spheroids.

We first introduce a system of spheroidal coordinates that are useful in
solving for the gravitational potential $\delta\Phi$.  Thus we introduce
the coordinates $(\mu,\zeta,\varphi)$ that are related to the usual
Cartesian coordinates $(x,y,z)$ by the transformation:

\beq
x=a \sqrt{(\zeta^2 + 1)(1-\mu^2)}\cos\varphi,\label{3.1}
\eeq

\beq
y=a \sqrt{(\zeta^2 + 1)(1-\mu^2)}\sin\varphi,\label{3.2}
\eeq

\beq
z=a \zeta\mu,\label{3.3}
\eeq

\noindent where $a$ is defined in Eq.~(\ref{2.3}) above.  These
are the standard oblate spheroidal coordinates.  It is straightforward
to show that

\beq
{x^2+y^2\over 1+\zeta^2} + {z^2\over \zeta^2} = a^2.
\label{3.3.1}
\eeq

\noindent Thus the surfaces of constant $\zeta$ are oblate spheroids,
with $\zeta=\zeta_o$ corresponding to the surface of the star.  The
coordinate $\zeta$ has the range $0\leq \zeta \leq \zeta_o$ within
the star, and $\zeta\geq\zeta_o$ outside.  The surface $\zeta=0$
corresponds to a disk of radius $a$ within the equatorial plane of the
star.  The nature of the surfaces of constant $\mu$ can similarly be
explored by noting that

\beq
{x^2+y^2\over 1-\mu^2} - {z^2\over \mu^2} = a^2.
\label{3.3.2}
\eeq

\noindent Thus the constant $\mu$ surfaces are hyperbolas.  The
coordinate $\mu$ is confined to the range $-1 \leq \mu \leq 1$, with
$\mu^2 = 1$ corresponding to the rotation axis of the star, and $\mu =
0$ to the portion of the equatorial plane outside the disk of radius
$a$.  The coordinate $\varphi$ measures angles about the rotation axis. 

The Equation~(\ref{1.7}) for the gravitational potential in these spheroidal
coordinates becomes

\beqa
{\partial\over\partial \zeta}\left[(\zeta^2+1){\partial\delta\Phi\over
\partial\zeta}\right]&&+
{\partial\over\partial \mu}\left[(1-\mu^2){\partial\delta\Phi\over
\partial\mu}\right]\nonumber\\&&+
{\zeta^2+\mu^2\over (\zeta^2+1)(1-\mu^2)}{\partial^2\delta\Phi\over
\partial\varphi^2}=0,\label{3.4} 
\eeqa

\noindent except on the surface of the star $\zeta=\zeta_o$.  The
separable solutions to these equations are functions of the form
$P_l^m(i\zeta)P_l^m(\mu)e^{im\varphi}$ and
$Q_l^m(i\zeta)P_l^m(\mu)e^{im\varphi}$.  The associated Legendre
functions $Q_l^m(\mu)$ diverge at $\mu^2 =1$, consequently only the
functions $P_l^m(\mu)$ appear in these solutions.  The associated
Legendre functions $P_l^m(i\zeta)$ diverge as $\zeta^l$ while the
$Q_l^m(i\zeta)$ vanish as $\zeta^{-(l+1)}$ in the limit
$\zeta\rightarrow \infty$.  Thus the gravitational potential in the
exterior of the star, $\zeta\geq \zeta_o$, must have the form:

\beq
\delta \Phi = \alpha {Q_l^m(i\zeta)\over Q_l^m(i\zeta_o)}P_l^m(\mu)
e^{im\varphi}.\label{3.5}
\eeq

\noindent for some constant $\alpha$.  In the interior of the star
$\zeta\leq \zeta_o$ the situation is more complicated.  Both
$P_l^m(i\zeta)$ and $Q_l^m(i\zeta)$ are bounded in the limit
$\zeta\rightarrow 0$.  However, we must insure that the solution is
smooth across the disk $\zeta=0$.  The functions $Q_l^m(i\zeta)$ are
non-zero at $i\zeta = 0$ (see Bateman \cite{bateman} Eqs.~3.4.9, 3.4.20
and 3.4.21).  For the case of $l+m$ odd the function
$Q_l^m(i\zeta)P_l^m(\mu)$ is therefore discontinuous at the disk
$\zeta=0$, and consequently it does not satisfy Laplace's equation
there.  Similarly for $l+m$ even the function $Q_l^m(i\zeta)P_l^m(\mu)$
is continuous but not differentiable at $\zeta=0$, and so it does not
satisfy Laplace's equation there either.  Thus, in the interior of the
star, $\zeta\leq\zeta_o$, the solution to Eq.~(\ref{1.7}) for given $l$
and $m$ is

\beq
\delta \Phi = \alpha {P_l^m(i\zeta)\over P_l^m(i\zeta_o)}P_l^m(\mu)
e^{im\varphi}.\label{3.6}
\eeq

\noindent  The potentials in Eqs.~(\ref{3.5}) and (\ref{3.6})
have been normalized so that $\delta\Phi$ is continuous at the
surface of the star $\zeta=\zeta_o$.

Following the analysis of the gravitational potential equation, we
introduce a second system of coordinates $(\xi,\tilde{\mu},\varphi)$
which allow the equation for the hydrodynamic potential,
Eq.~(\ref{1.6}), to be written in a convenient form.  These coordinates
are related but not identical to those used by Bryan~\cite{bryan}:

\beq x=b\sqrt{(1-\xi^2)(1-\tilde{\mu}^2)}\cos\varphi,\label{3.8} \eeq

\beq
y=b\sqrt{(1-\xi^2)(1-\tilde{\mu}^2)}\sin\varphi,\label{3.9}
\eeq

\beq
z=b\xi\tilde{\mu}{\sqrt{4-\kappa^2}\over\kappa}.\label{3.10}
\eeq

\noindent Here the focal radius $b$ is related to $a$ by

\beq
b^2={a^2\over 4-\kappa^2}\left[4(1+\zeta_o^2)-\kappa^2
\right].\label{3.11}
\eeq

\noindent The parameter $b$ is real for frequencies, such as those of
the $r-$modes, which satisfy $\kappa^2< 4$.  As we shall see, the
Eq.~(\ref{1.6}) for the potential $\delta U$ separates nicely in terms
of these coordinates.  However these new coordinates are rather unusual,
so we present an in depth discussion of them in the Appendix.  In
summary, the coordinates $\xi$ and $\tilde{\mu}$ cover the interior
of the star when their values are confined to the domains $\xi_o\leq
\xi\leq 1$ and $-\xi_o\leq\tilde{\mu}\leq\xi_o$, where $\xi_o$
is defined as

\beq
\xi_o^2= {a^2\zeta_o^2\over b^2}{\kappa^2\over 4-\kappa^2}
={\zeta_o^2\kappa^2\over 4(1+\zeta_o^2)-\kappa^2}<{\kappa^2\over 4}
<1.\label{3.12}
\eeq

\noindent The surface $\xi=1$ corresponds to the rotation axis of the
star, and the surface $\tilde{\mu}=0$ corresponds to the equatorial
plane of the star.  The surface of the star, $\zeta=\zeta_o$, is divided
into three regions in this coordinate system.  The portions of the
stellar surface nearest the two branches of the rotation axis correspond
to the surfaces $\tilde{\mu}=\pm\xi_o$, while the portion of the stellar
surface that includes the equator corresponds to the surface
$\xi=\xi_o$.  The coordinate $\tilde{\mu}$ coincides with the value of
the coordinate $\mu$ in that portion of the surface of the star where
$\xi=\xi_o$.  In the other portions of the surface of the star the value
of $\mu$ coincides with $\pm\xi$.  These facts will be essential in
imposing the boundary conditions in the next section. 

The Equation~(\ref{1.6}) for the potential $\delta U$ reduces to the
following in terms of the coordinates $(\xi,\tilde{\mu},\varphi)$

\beqa
{\partial\over\partial \xi}\left[(\xi^2-1){\partial\delta U\over
\partial\xi}\right]&&+
{\partial\over\partial \tilde{\mu}}\left[(1-\tilde{\mu}^2)
{\partial\delta U\over
\partial\tilde{\mu}}\right]\nonumber\\&&+
{\xi^2-\tilde{\mu}^2\over (\xi^2-1)(1-\mu^2)}{\partial^2\delta U\over
\partial\varphi^2}=0.\label{3.13} 
\eeqa

\noindent The separated solutions of this equation are associated
Legendre functions of $\xi$ and $\tilde{\mu}$.  The coordinate $\xi$
includes $\xi=1$ in its range, so the non-singular separated solutions
to Eq.~(\ref{3.13}) are $P_l^m(\xi)P_l^m(\tilde{\mu})e^{im\varphi}$ and
$P_l^m(\xi)Q_l^m(\tilde{\mu})e^{im\varphi}$.  The ``angular'' coordinate
$\tilde{\mu}$ does not include $\pm 1$ in its range.  Thus at this
stage, it is not possible to eliminate the $Q_l^m(\tilde{\mu})$ solution
without imposing the boundary conditions. 

\section{Imposing The Boundary Conditions}
\label{sectionIV}

In order to obtain the physical solutions to the stellar pulsation
equations, we must now impose the boundary conditions, Eqs.~(\ref{1.10})
and (\ref{1.11}).  The simplest boundary condition is the one that
involves the derivatives of $\delta\Phi$, Eq.~(\ref{1.11}).  In the
coordinates $(\mu,\zeta,\varphi)$ used to find the solution for $\delta
\Phi$ in Eqs.~(\ref{3.5}) and (\ref{3.6}), the unit normal vector to the
surface of the spheroid $n^a$ has only one nonvanishing component,
$n^\zeta$:

\beq
n^\zeta = {1\over a}\sqrt{\zeta_o^2+1\over \zeta_o^2+\mu^2}.
\label{4.1}
\eeq

\noindent Thus, the normal derivatives that appear in the boundary
condition, $n^a\nabla_a\delta\Phi$, can be expressed simply as
$n^\zeta\partial_\zeta\delta\Phi$.  The gradient of the pressure,
$n^a\nabla_ap$ that appears in Eq.~(\ref{1.11}) is given by
Eq.~(\ref{2.4}).  When evaluated at the surface of the spheroid
this reduces to

\beqa
n^a\nabla_ap=-4\pi G\rho^2 a\,\zeta_o&&\sqrt{(\zeta_o^2+1) (\zeta_o^2+\mu^2)}
\nonumber \\&&\qquad\times(1-\zeta_o\cot^{-1}\zeta_o).\label{4.2}
\eeqa

\noindent Thus, using Eqs.~(\ref{3.5}), (\ref{3.6}), (\ref{4.1}) and
(\ref{4.2}) the boundary condition Eq.~(\ref{1.11}) on $\delta\Phi$ is
equivalent to

\beqa
\alpha{\zeta_o^2+1\over Q_l^m(i\zeta_o)}&&{dQ_l^m(i\zeta_o)\over d\zeta}
P_l^m(\mu) e^{im\varphi}=\nonumber\\
&&\,\,\,
\alpha{\zeta_o^2+1\over P_l^m(i\zeta_o)}{dP_l^m(i\zeta_o)\over d\zeta}
P_l^m(\mu) e^{im\varphi}\nonumber\\
&&\,\,\,-{\alpha P_l^m(\mu) e^{im\varphi}
+\delta U\over \zeta_o(1-\zeta_o\cot^{-1}\zeta_o)}.
\label{4.3}
\eeqa

\noindent The first immediate consequence of this boundary condition is
that the potential $\delta U$ must be proportional to $P_l^m(\mu)$ on
the surface of the star.  In the last section we found that the
potential $\delta U$ was some linear combination of
$P_l^m(\xi)P_l^m(\tilde{\mu})e^{im\varphi}$ and
$P_l^m(\xi)Q_l^m(\tilde{\mu})e^{im\varphi}$.  As we show in the
Appendix, the surface of the star is somewhat complicated in the $(\xi,
\tilde{\mu},\varphi)$ coordinate system.  For the portion of the surface
of the star that includes the equator, we found that $\xi=\xi_o$ and
$\tilde{\mu}=\mu$.  This fixes the angular dependence of $\delta U$. 
Therefore, throughout the star $\delta U$ must have the form

\beq
\delta U = \beta {P_l^m(\xi)\over P_l^m(\xi_o)}P_l^m(\tilde{\mu})
e^{im\varphi},
\label{4.4}
\eeq

\noindent where $\beta$ is an arbitrary constant.  On the portion of the
surface of the star that includes the equator, this expression reduces
to $\delta U = \beta P_l^m(\tilde{\mu}) e^{im\varphi} = \beta P_l^m(\mu)
e^{im\varphi}$.  On the portion of the surface that includes the
positive rotation axis, $\tilde{\mu}=\xi_o$, the expression for $\delta
U$ reduces to $\delta U = \beta P_l^m(\xi) e^{im\varphi} = \beta
P_l^m(\mu)e^{im\varphi}$ since $\xi=\mu$ here.  Finally, on the portion
of the surface that includes the negative rotation axis,
$\tilde{\mu}=-\xi_o$, the expression for $\delta U$ also reduces to
$\delta U = \beta P_l^m(-\xi) e^{im\varphi} = \beta
P_l^m(\mu)e^{im\varphi}$ since $\xi=-\mu$ 
here.  Consequently, the potential $\delta U$ reduces to the expression
$\delta U = \beta P_l^m(\mu) e^{im\varphi}$ everywhere on the surface of
the star.  Thus, the boundary condition on $\delta\Phi$ reduces to

\beqa
\alpha{\zeta_o^2+1\over Q_l^m(i\zeta_o)}{dQ_l^m(i\zeta_o)\over d\zeta}
=&&
\alpha{\zeta_o^2+1\over P_l^m(i\zeta_o)}{dP_l^m(i\zeta_o)\over d\zeta}
 \nonumber\\
&&\quad-{\alpha +\beta\over \zeta_o(1-\zeta_o\cot^{-1}\zeta_o)}.
\label{4.5}
\eeqa

 We now see why it was necessary to obtain the solutions for $\delta U$
in the strange and complicated $(\xi,\tilde{\mu},\varphi)$ coordinate
system.  These unusual coordinates have two essential properties: first
they allow the solutions to Eq.~(\ref{1.6}) to be found in separated
form, and second they have the property that one of the coordinates
reduces on the surface of the star to the angular coordinate $\mu$. 
This last property was needed to allow us to satisfy the boundary
conditions using simple separated solutions for both $\delta U$
and $\delta \Phi$. 

The boundary condition, Eq.~(\ref{1.10}), on the potential $\delta U$
is unfortunately somewhat more complicated.  The tensor $Q^{ab}$ that is
used in Eq.~(\ref{1.10}) is most simply expressed in cylindrical
coordinates (see Eq.~\ref{1.3}).  Therefore it is simplest to consider
the boundary conditions on $\delta U$ in these coordinates.  Let
$\varpi^2=x^2+y^2$ denote the cylindrical radial coordinate.  Then, the
boundary condition Eq.~(\ref{1.10}) can be expressed as

\beqa
(\kappa^2-4)n^z\partial_z\delta U&&+\kappa^2n^\varpi\partial_\varpi\delta U
+{2m\kappa\over \varpi}n^\varpi \delta U =\nonumber \\
&&-\kappa^2(\kappa^2-4)\Omega^2\rho
{\delta U+\delta\Phi\over n^a\nabla_ap}.\label{4.6}
\eeqa

\noindent The components of the unit normal vector to the surface of the
spheroid, $n^a$, that appear in Eq.~(\ref{4.6}) can be obtained by
taking the gradient of the function that appears on the left side of
Eq.~(\ref{2.3}):

\beq
n^\varpi=\sqrt{\zeta_o^2
(1-{\mu}^2)\over  \zeta_o^2+{\mu}^2},\label{4.7}
\eeq

\beq
n^z={\mu}\sqrt{\zeta_o^2+1\over \zeta_o^2
+{\mu}^2}.\label{4.8}
\eeq

\noindent The partial derivatives $\partial_z\delta U$ and
$\partial_\varpi\delta U$ that appear in Eq.~(\ref{4.6}) are more
difficult to evaluate.  To do this we must evaluate the Jacobian matrix
that determines the coordinate transformation defined in
Eqs.~(\ref{3.8}) through (\ref{3.10}).  The needed partial derivatives are
given in the Appendix as Eqs.~(\ref{a11}) through (\ref{a14}).  These
expressions can now be used to transform the derivatives $\partial_z$
and $\partial_\varpi$ needed in the boundary condition into expressions
in terms of $\partial_\xi$ and $\partial_{\tilde{\mu}}$.  When evaluated
on the portion of the surface of the star with $\xi=\xi_o$ and
$\tilde{\mu}=\mu$ (using the relationships in Eqs.~\ref{3.11} and
\ref{3.12}), we obtain the following

\beqa
(\kappa^2-4)&&n^z\partial_z\delta U 
+ \kappa^2 n^\varpi\partial_\varpi\delta U \nonumber\\
&&=-{\kappa( 4-\kappa^2)\over a\sqrt{4(\zeta_o^2+1)-\kappa^2}
}\sqrt{\zeta_o^2+1\over \zeta_o^2+\mu^2}
\partial_\xi\delta U.
\label{4.9}
\eeqa

\noindent Similarly, when evaluated on the portions of the surface
with $\tilde{\mu}=\pm\xi_o$ and $\xi=\pm\mu$ we find

\beqa
(\kappa^2-4)&&n^z\partial_z\delta U 
+ \kappa^2 n^\varpi\partial_\varpi\delta U \nonumber\\
&&=\mp{\kappa( 4-\kappa^2)\over a\sqrt{4(\zeta_o^2+1)-\kappa^2}
}\sqrt{\zeta_o^2+1\over \zeta_o^2+\mu^2}
\partial_{\tilde{\mu}}\delta U.
\label{4.10}
\eeqa

\noindent Finally then, we may combine the results of Eqs.~(\ref{3.5}),
(\ref{4.4}), (\ref{4.9}), and (\ref{4.10}) to obtain the following
representation of the boundary condition Eq.~(\ref{4.6}):

\beqa
&&{\zeta_o^2+1\over P_l^m(\xi_o)\sqrt{4(\zeta_o^2+1)-\kappa^2}}
{dP_l^m(\xi_o)\over d\xi}\beta-{2m\zeta_o\over 4-\kappa^2}\beta 
\nonumber\\&&\qquad\qquad=
{(1+3\zeta_o^2)\cot^{-1}\zeta_o-3\zeta_o\over
2(1-\zeta_o\cot^{-1}\zeta_o)}\kappa(\alpha+\beta).
\label{4.11}
\eeqa

\noindent We point out that Eq.~(\ref{4.11}) is valid on the entire
surface of the star.  For the portion of the surface where
$\tilde{\mu}=-\xi_o$ and $\xi=-\mu$ on the surface, it is helpful to
remember that $P_l^m(-\xi_o)P_l^m(-\mu)= P_l^m(\xi_o)P_l^m(\mu)$, while
$P_l^m(-\mu)\,dP_l^m(-\xi_o)/d\xi = -P_l^m(\mu)\,
dP_l^m(\xi_o)/d\xi$. 

In summary then the boundary conditions, Eqs.~(\ref{4.5}) and (\ref{4.11})
are given by

\beq B(\zeta_o) \alpha = -{\alpha + \beta\over
\zeta_o(1-\zeta_o\cot^{-1}\zeta_o)}, \label{4.12} \eeq

\beq
A(\kappa,\zeta_o)\beta = 
{(1+3\zeta_o^2)\cot^{-1}\zeta_o-3\zeta_o\over
2(1-\zeta_o\cot^{-1}\zeta_o)}\kappa(\alpha+\beta),
\label{4.13}
\eeq

\noindent where $A(\kappa,\zeta_o)$ and $B(\zeta_o)$ are defined
by

\beq
A(\kappa,\zeta_o)
={\zeta_o^2+1\over P_l^m(\xi_o)\sqrt{4(\zeta_o^2+1)-\kappa^2}
}{dP_l^m(\xi_o)\over d\xi}
-{2m\zeta_o\over 4-\kappa^2},
\label{4.14}
\eeq

\beq
B(\zeta_o)=
{\zeta_o^2+1\over Q_l^m(i\zeta_o)}{dQ_l^m(i\zeta_o)\over d\zeta}
-{\zeta_o^2+1\over P_l^m(i\zeta_o)}{dP_l^m(i\zeta_o)\over d\zeta}.
\label{4.15}
\eeq

\noindent Note that $\xi_o$ that appears on the right side of
Eq.~(\ref{4.14}) is an implicit function of $\kappa$ and $\zeta_o$ as
given in Eq.~(\ref{3.12}). 

\section{The Eigenvalues} \label{sectionV}

The boundary conditions Eqs.~(\ref{4.12}) and (\ref{4.13}) can be
satisfied for only certain values of the eigenvalue $\kappa$.  It is
easy to see that the necessary and sufficient condition that there
exists a solution to the boundary conditions is

\beqa
0=F(&&\kappa,\zeta_o)\equiv
A(\kappa,\zeta_o)B(\zeta_o)+\nonumber\\&&{2 A(\kappa,\zeta_o)-\kappa\zeta_o 
B(\zeta_o)\bigl[(1+3\zeta_o^2)\cot^{-1}\zeta_o-3\zeta_o\bigr]
\over 2\zeta_o(1-\zeta_o\cot^{-1}\zeta_o)}.\nonumber\\
\label{5.3}
\eeqa

\noindent We have verified that Eq.~(\ref{5.3}) is exactly equivalent
(i.e.  up to changes in notation) to Eq.~(60) of Bryan~\cite{bryan}. 

We now wish to evaluate this eigenvalue equation analytically for the
generalized $r-$modes of slowly rotating stars.  The parameter
$\zeta_o$ determines the angular velocity of the star through
Eq.~(\ref{2.1}).  Large values of $\zeta_o$ correspond to small
angular velocities, and so we may expand our equations in inverse
powers of $\zeta_o$.  The leading order terms in such an expression
for the angular velocity are

\beq
{\Omega^2\over \pi G \rho} = {8\over 15\zeta_o^2}\left[
1-{6\over 7\zeta_o^2} + {\cal O}(\zeta_o^{-4})\right].
\label{5.4}
\eeq

We now wish to obtain solutions to the eigenvalue equation 
$F(\kappa,\zeta_o)=0$ for large values of $\zeta_o$.  We do
this by expanding the expressions on the right side of
Eq.~(\ref{5.3}) in inverse powers of $\zeta_o$:

\beqa
F(\kappa,\zeta_o)=-(l-1)\zeta_o^2&&
\Biggl[{1\over P_l^m(\kappa/2)}{dP_l^m(\kappa/2)\over d\xi}-
{4m\over 4-\kappa^2}\Biggr]\nonumber\\&& 
\times[1+ {\cal O}(\zeta_o^{-2})].
\label{5.10}
\eeqa

\noindent Setting the coefficient of this lowest-order term to zero,
we obtain an equation for the lowest-order expression for the
eigenvalue, $\kappa_o$:

\beq
{dP_l^m(\kappa_o/2)\over d\xi}=
{4m\over 4-\kappa_o^2}P_l^m(\kappa_o/2).
\label{5.11}
\eeq

\noindent Using the Rodrigues formula for the associated Legendre
functions, this equation can be transformed (for the $m\geq0$ modes)
into the form

\beq
0=m{d^{\,m}P_l(\kappa_o/2)\over d\xi^m} + \left({\kappa_o\over 2}-1\right)
{d^{\,m+1}P_l(\kappa_o/2)\over d\xi^{m+1}},\label{5.111}
\eeq

\noindent where $P_l$ is the Legendre polynomial of degree $l$.
Equation~(\ref{5.111}) is equivalent to Bryan's Eq.~(83)~\cite{bryan}.
This equation admits $l-m$ (or $l-1$ for the $m=0$ case) distinct
roots all of which lie in the interval $-2<\kappa_o<2$~\cite{note2}.
For the case $l=m+1$ the single root of Eq.~(\ref{5.111}) is
$\kappa_o=2/(m+1)$.  This agrees with the frequency of the classical
$r$-mode of order $m$ as found for example by Papaloizou and
Pringle~\cite{p&p}.  The modes with $m<0$ are equivalent to those with
$m>0$: if $\kappa_o$ is a solution to Eq.~(\ref{5.11}) for some $m$,
then $-\kappa_o$ is a solution for $-m$.  In Table~\ref{table1} we
present numerical solutions to Eq.~(\ref{5.11}) for a range of
different values of $l$ and $m$.  We see that for each value of $m$
there exist solutions of this equation for each value of $l\geq m+1$.
For each value of $l$ and $m$ there are $l-m$ different solutions.
Thus, there exist a vast number of modes whose frequencies vanish
linearly with the angular velocity of the star.  We indicate with a
$*$ those frequencies in Table~\ref{table1} that satisfy the condition
$\omega(\omega+m\Omega)<0$.  The modes satisfying this condition would
be driven unstable by gravitational radiation reaction in the absence
of internal fluid dissipation (i.e.  viscosity)~\cite{LOM}.

\begin{table}[top] \caption{The eigenvalues $\kappa_o$ of the $r$-modes
of the Maclaurin spheroids in the limit of low angular velocities.  The
frequencies of these modes are related to $\kappa_o$ by
$\omega=(\kappa_o-m)\Omega$ in the low angular velocity limit.  Those
frequencies denoted with $*$ have the property that
$\omega(\omega+m\Omega)<0$.  These $*$ modes are subject to a
gravitational radiation driven secular instability.\label{table1}}
\begin{tabular}{clllll} &$m=0$ &$m=1$ &$m=2$ &$m=3$ &$m=4$\\ \tableline
$l=2$ & $\,\,$0.0000 & $\,\,$1.0000\hfill \\ $l=3$ & $\,\,$0.8944 &
$\,\,$1.5099 & $\,\,$0.6667$*$ \\
      &-0.8944 &-0.1766 \\
$l=4$ & $\,\,$1.3093 & $\,\,$1.7080 & $\,\,$1.2319$*$ 
         & $\,\,$0.5000$*$ \\
      & $\,\,$0.0000 & $\,\,$0.6120$*$ & -0.2319 \\
      &-1.3093 &-0.8200 \\
$l=5$ & $\,\,$1.5301 & $\,\,$1.8060 &$\,\,$1.4964$*$ 
      & $\,\,$1.0532$*$ & $\,\,$0.4000$*$ \\
      & $\,\,$0.5705 & $\,\,$1.0456 & $\,\,$0.4669$*$ 
      & -0.2532 \\
      & -0.5795 & -0.0682 & -0.7633 \\
      & -1.5301 & -1.1834 \\
$l=6$ & $\,\,$1.6604 & $\,\,$1.8617 & $\,\,$1.6434$*$ 
      & $\,\,$1.3402$*$ & $\,\,$0.9279$*$ \\
      & $\,\,$0.9377 & $\,\,$1.3061 & $\,\,$0.8842$*$ 
      & $\,\,$0.3779$*$ & -0.2613 \\
      & $\,\,$0.0000 & $\,\,$0.4405$*$ & -0.1018 
      & -0.7181 \\
      & -0.9377 & -0.5373 & -1.0926 \\
      &-1.6604 & -1.4042 \\
\end{tabular}
\end{table}
\nobreak

Next we wish to extend the formula for the eigenvalue to higher
angular velocity.  We define the next term in the expansion
of the eigenvalue as

\beq
\kappa = \kappa_o + \kappa_2 \zeta_o^{-2} + {\cal O}(\zeta_o^{-4}).
\label{5.12}
\eeq

\noindent Using this definition and Eq.~(\ref{5.11}) for $\kappa_o$
we find the next order term in $F(\kappa,\zeta_o)$ to be

\beqa
F(\kappa,\zeta_o)=-(l-1)
&&\left[
{m\over 2} + {\kappa_ol(l+1)\over 4} 
- {2\kappa_2l(l+1)\over 4-\kappa_o^2}\right]\nonumber\\
&& +\case{2}{5}(2l+1)\kappa_o + {\cal O}(\zeta_o^{-2}).
\label{5.14}
\eeqa

\noindent We can now determine the second order correction to the 
eigenvalue of the mode by solving $F(\kappa,\zeta_o)=0$ for
$\kappa_2$:

\beq
\kappa_2 = {\kappa_o(4-\kappa_o^2)\over 2l(l+1)}
\left[ {m\over 2\kappa_o} + {l(l+1)\over 4}
-{2\over 5} {2l+1\over l-1}\right].
\label{5.15}
\eeq

\noindent  In the case of the classical $r$-modes with $l=m+1$,
this expression reduces to:

\beq
\kappa_2 = {4m\over (m+1)^4}\left[{(m+1)^2\over 2}
- {2\over 5} {2m+3\over m}\right].
\label{5.16}
\eeq

\noindent Thus, for the classical $r-$modes the frequency of the
mode is given by~\cite{note1}

\beqa
\omega+m\Omega = \Omega\Biggl\{&&{2\over m+1}\nonumber \\
&&+{3m\over (m+1)^2}
\left[\frac{5}{4}-{2m+3\over m(m+1)^2}\right]{\Omega^2\over
\pi G \rho}\Biggr\}\nonumber\\
&&+{\cal O}(\Omega^5).
\label{5.16.1}
\eeqa

For rapidly rotating stars we must solve the eigenvalue
Eq.~(\ref{5.3}) numerically.  This presents certain unusual numerical
difficulties.  In particular the associated Legendre functions
$P_l^m(i\zeta)$ and $Q_l^m(i\zeta)$ that appear in Eq.~(\ref{4.15}) are
not commonly encountered.  Thus, we digress briefly here to describe
how these functions may be evaluated numerically.  First we note that
these functions are essentially real.  In particular the functions
$\tilde{P}_l^m(\zeta)$ and $\tilde{Q}_l^m(\zeta)$ defined by

\beq
P_l^m(i\zeta)=i^{\,l}\tilde{P}_l^m(\zeta),\label{5.17}
\eeq

\beq
Q_l^m(i\zeta)=i^{\,l+1}\tilde{Q}_l^m(\zeta),\label{5.18}
\eeq

\noindent are real for real values of $\zeta$.  The functions
$\tilde{P}_l^m(\zeta)$ can be evaluated numerically using essentially
the same algorithms used to evaluate their counterparts on the real
axis.  In particular

\beq
\tilde{P}_m^m(\zeta)=(2m-1)!!(\zeta^2+1)^{m/2},\label{5.19}
\eeq

\beq
\tilde{P}_{m+1}^m(\zeta)=(2m+1)\zeta\tilde{P}_m^m(\zeta)
.\label{5.20}
\eeq

\noindent These expressions can be used as initial values for the
recursion formula,

\beq
(l-m)\tilde{P}_l^m(\zeta)=(2l-1)\zeta\tilde{P}_{l-1}^m(\zeta)
+(l+m-1)\tilde{P}_{l-2}^m(\zeta),\label{5.21}
\eeq

\noindent from which the higher order functions can be determined.
This approach does not work for the $\tilde{Q}_l^m$ however.  The problem
is that we need to evaluate these functions over a fairly wide
range of their arguments, e.g. $0.25\lesssim \zeta \lesssim 75$.  For large
values of $\zeta$ the recursion for the $\tilde{Q}_l^m$ involves
a high degree of cancellation among the various terms.  Standard
computers simply do not have the numerical precision to
perform these calculations to sufficient accuracy.  Instead we
rely on an integral representation of $\tilde{Q}_l^m$.  Based
on Bateman's Eq.~(3.7.5) \cite{bateman}, we find

\beqa
\tilde{Q}_l^m(\zeta)= &&{(-1)^{l+m+1}(l+m)!\over 2^{l+1}l!(\zeta^2+1)^{m/2}}
\nonumber\\
&&\quad\times\int_{-1}^1(\zeta^2+t^2)^{(m-l-1)/2}(1-t^2)^l\nonumber\\
&&\qquad\quad\times\cos\bigl[(l+1-m)\tan^{-1}(t/\zeta)\bigr]dt.\label{5.22}
\eeqa

\noindent The integrand in this expression is well behaved for all
values of $\zeta>0$, and the integrals may be determined numerically
quite easily.

\bfig[t] \centerline{\psfig{file=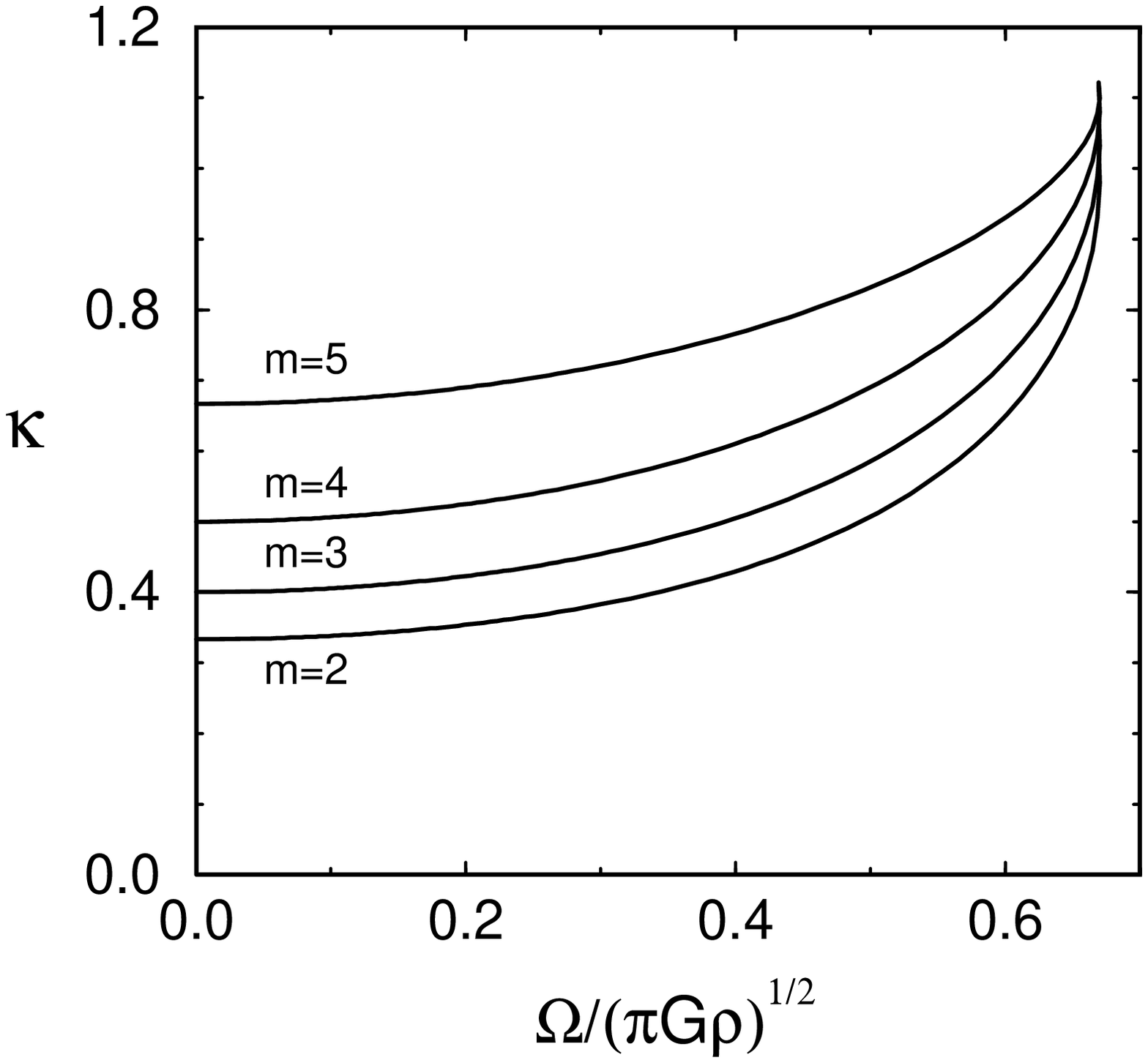,height=2.0in}} \vskip 0.3cm
\caption{Angular velocity dependence of the eigenvalues $\kappa$ of the
classical $r-$modes, i.e.  those modes with $l=m+1$ for $l\leq 6$..  The
frequencies of these modes are related to $\kappa$ by
$\omega=(\kappa-m)\Omega$.  \label{fig1}} \efig

Using these numerical techniques, then, it is straightforward to solve
for the eigenvalues of Eq.~(\ref{5.3}), $F(\kappa,\zeta_o)=0$, over the
relevant range of the parameter $0.25\leq \zeta_o \leq 75$.
Figures~\ref{fig1}, \ref{fig2}, and \ref{fig3} display the angular
velocity dependence of the eigenvalue $\kappa$ for a number of
$r-$modes.  Figure~\ref{fig1} depicts $\kappa$ for the `classical'
$r-$modes, $l=m+1$, with $l\leq 6$.  This $l=m+1=3$ mode is the one
found by Lindblom, Owen, and Morsink \cite{LOM} to be sufficiently
unstable due to gravitational radiation emission that it is expected
to cause all hot young rapidly rotating neutron stars to spin down to
low angular velocities within about one year.  We have verified that
our numerical solutions of Eq.~(\ref{5.3}) agree with those
of Eq.~(\ref{5.16.1}), up to terms that scale as $\Omega^5$. 

Figure~\ref{fig2} presents the angular velocity dependence of the
frequency of the classical $l=m+1=3$ $r-$mode as measured in an inertial
frame.  The solid curve corresponds to the exact solution to
Eq.~(\ref{5.3}), while the dot-dashed and dashed curves represent the
first and second order approximations (respectively) given in
Eq.~(\ref{5.16.1}).  The units in Fig.~\ref{fig2} for both the vertical
and horizontal axis scale as $\sqrt{\pi G \rho}$.  The value of
$\rho=7\times 10^{14}$gm/cm${}^3$ chosen here represents a typical average
density for a neutron star.  Figure~\ref{fig2} illustrates three
interesting features about the frequencies of this mode in large angular
velocity stars.  First, the frequency is only about $2/3$ that predicted
by the first-order formula for stars rotating near their maximum angular
velocity.  This means that the gravitational radiation reaction force,
which scales as $(\omega+m\Omega)\omega^{5}$, could be about $1/5$ of
that predicted by the first order formula.  (Unless the mass and current
multipoles of the rapidly rotating models are much larger than their
non-rotating values.) Second, the frequencies of these modes first
increase and then decrease as the angular velocity of the star is
reduced.  This means that the time evolution of the gravitational
radiation signal from these sources will be more complicated than had
been anticipated by Owen, et al.~\cite{owen} on the basis of the first
order expression for the frequency.  In particular it appears that the
evolution of the frequency will not be monotonic for the most rapidly
rotating stars.  Third, the accuracy of the second-order formula for the
frequency is in fact considerably worse than that of the simple
first-order formula for very rapidly rotating stars.  This suggests that
any application of low angular velocity expansions of the $r-$modes to
the study of rapidly rotating stars is somewhat suspect. 

\bfig[t]
\centerline{\psfig{file=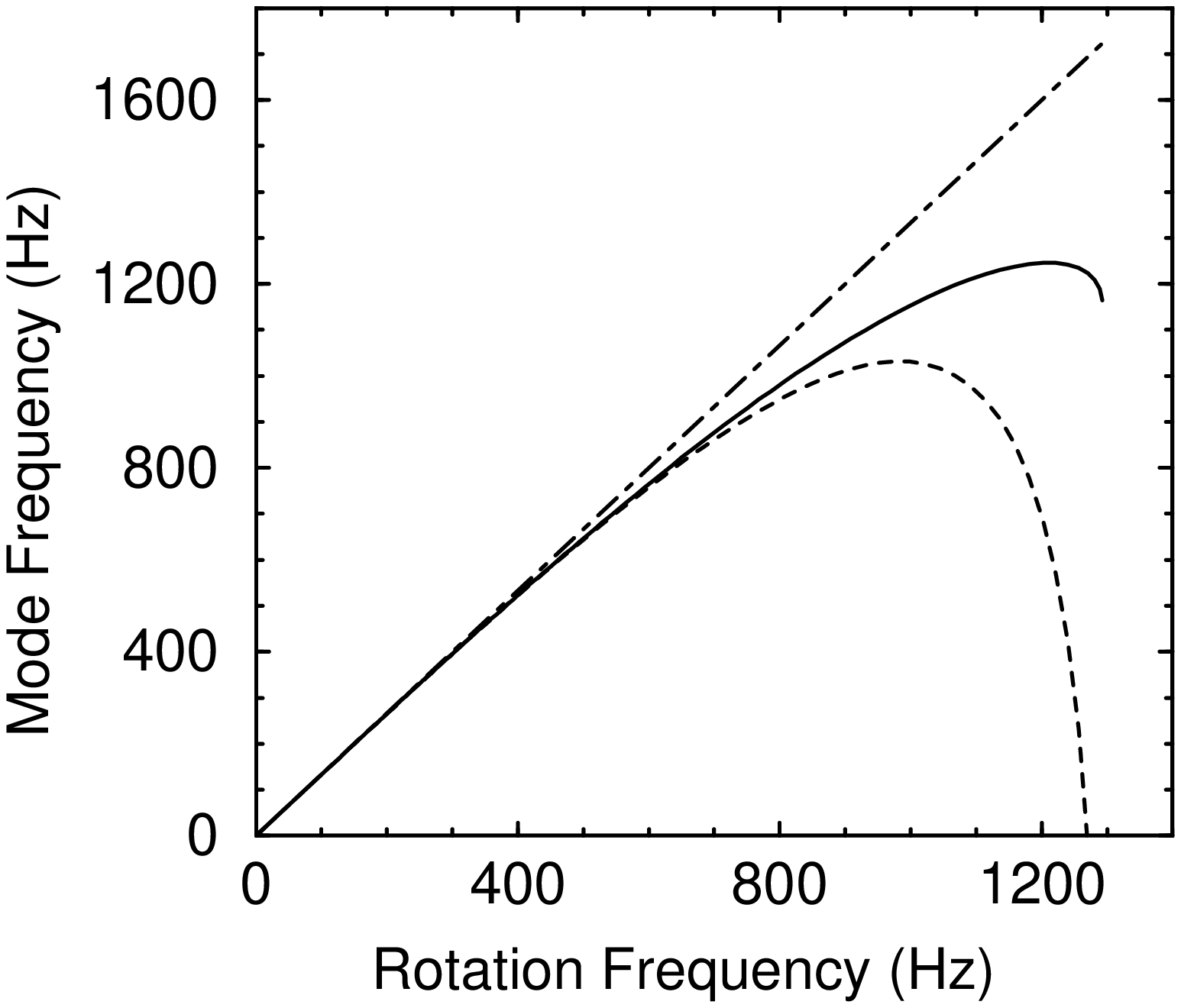,height=1.8in}} \vskip 0.3cm
\caption{Angular velocity dependence of the frequency
$\omega=(\kappa-m)\Omega$ of the classical $m=2$ $r-$mode.  The
solid curve gives $\omega$ corresponding to the exact solution
of Eq.~(\ref{5.3}), while the dot-dashed and dashed curves correspond
to the first and second order approximations respectively from
Eq.~(\ref{5.16.1}). \label{fig2}}
\efig

\bfig[t]
\centerline{\psfig{file=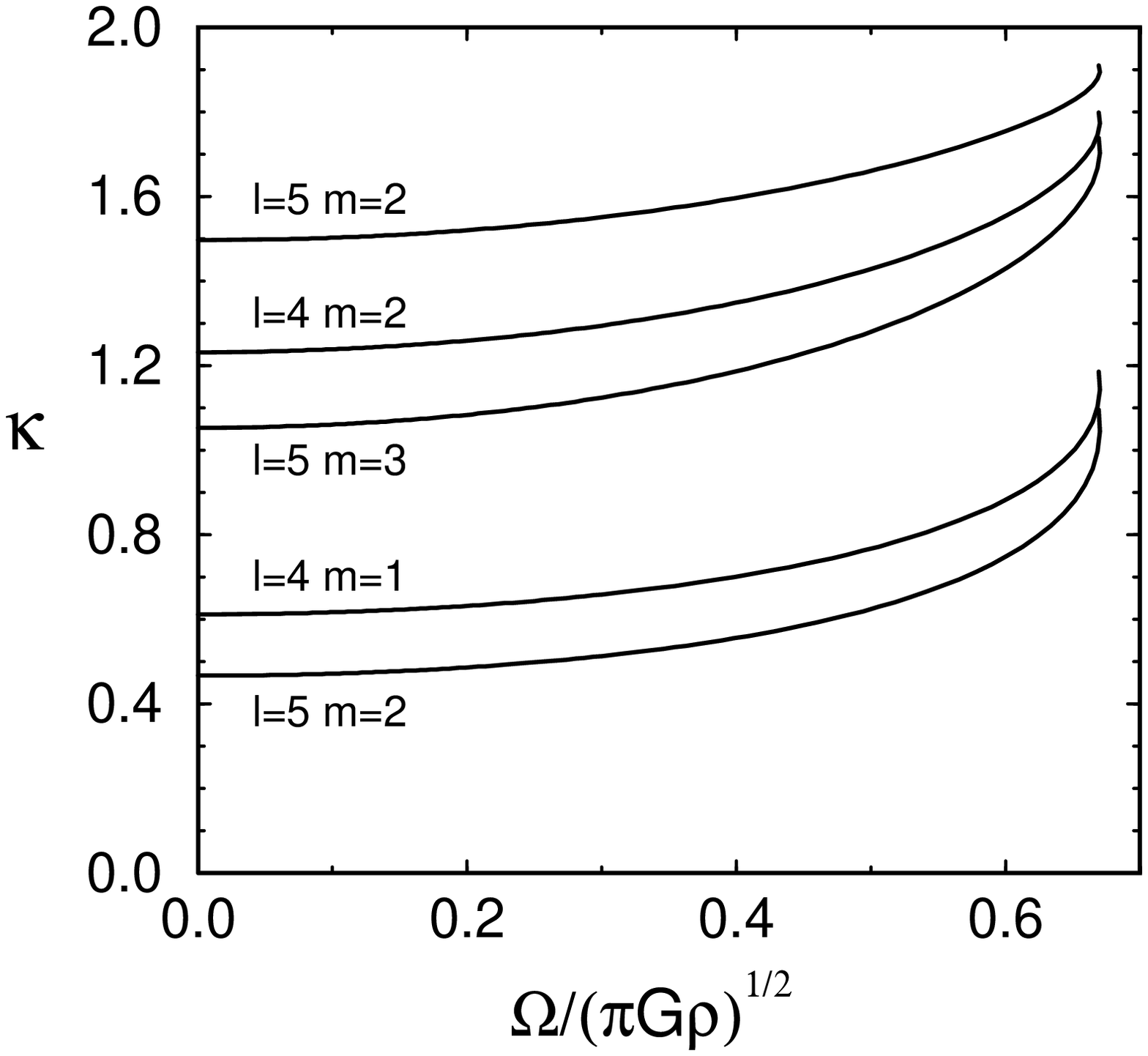,height=1.8in}}
\vskip 0.3cm
\caption{Angular velocity dependence of the eigenvalues $\kappa$
of $r-$modes with $5\geq l>m+1$ which are unstable to the
gravitational radiation instability.  The frequencies of these
modes are related to $\kappa$ by $\omega=(\kappa-m)\Omega$.
\label{fig3}}
\efig

Figure~\ref{fig3} depicts the angular velocity dependence of $\kappa$
for several other previously unstudied $r-$modes.  The modes depicted in
Fig.~\ref{fig3} all have the property that $\omega(\omega+m\Omega)<0$,
and hence these modes would all be subject to the gravitational
radiation secular instability in the absence of internal fluid
dissipation (i.e.  viscosity).  These additional modes all couple to
higher order gravitational moments than the classical $l=m+1=3$ mode. 
Thus, these additional modes probably do not play a significant role in
the astrophysical process which spins down hot young neutron stars.

\section{The Eigenfunctions}
\label{sectionVI}

The eigenfunctions associated with these modes are those given in
Eqs.~(\ref{3.5}), (\ref{3.6}), and (\ref{4.4}).  Inside the fluid we
have

\beq
\delta \Phi = \alpha {P_l^m(i\zeta)\over P_l^m(i\zeta_o)}P_l^m(\mu)
e^{im\varphi}e^{i(\kappa-m)\Omega t},\label{6.1}
\eeq

\beq
\delta U = \beta {P_l^m(\xi)\over P_l^m(\xi_o)}P_l^m(\tilde{\mu})
e^{im\varphi}e^{i(\kappa-m)\Omega t},
\label{6.2}
\eeq

\noindent while outside we have

\beq
\delta \Phi = \alpha {Q_l^m(i\zeta)\over Q_l^m(i\zeta_o)}P_l^m(\mu)
e^{im\varphi}e^{i(\kappa-m)\Omega t}.\label{6.3}
\eeq

\noindent The coefficients $\alpha$ and $\beta$ are determined by
solving the boundary conditions.  The eigenvalue Eq.~(\ref{5.3})
is the consistency condition for the existence of these solutions.
Given a solution of the eigenvalue equation then, the ratio of
$\alpha$ and $\beta$ can be determined from Eq.~(\ref{4.12}):

\beq
\beta=-\bigl[1+\zeta_o(1-\zeta_o\cot^{-1}\zeta_o)B(\zeta_o)\bigr]
\alpha.\label{6.4}
\eeq

\noindent It might appear at first glance that the eigenfunctions
associated with the $l-m$ distinct $r-$modes are identical.  However,
the coordinates $\xi$ and $\tilde{\mu}$ depend on the eigenvalue
$\kappa$.  Thus, the spatial dependence of $\delta U$ will be different
for each of these modes.  Interestingly enough, however, the spatial
dependence of the gravitational potential, $\delta\Phi$, depends only on
$l$ and $m$, and hence is the same for all $l-m$ distinct modes. 

While the expressions for the eigenfunctions are quite simple in terms
of the special spheroidal coordinates used here, they are rather
complicated when expressed in more traditional coordinates.  One
exception to this is the case of the classical $r-$modes, i.e.  those
with $l=m+1$.  In this case the needed associated Legendre function,
$P_{m+1}^m$, has the simple expression

\beq
P_{m+1}^m(\zeta)=(-1)^m (2m+1)!! \zeta (1-\zeta^2)^{m/2}.
\label{6.5}
\eeq

\noindent Using the expressions for the bi-spheroidal coordinates
$(\xi,\tilde{\mu})$ in terms of the cylindrical coordinates $(\varpi,
z)$ given in Eqs.~(\ref{a4}) and (\ref{a5}), it follows that

\beqa
P_{m+1}^m(\xi)&&P_{m+1}^m(\tilde{\mu})\nonumber\\
&&=(-1)^m(2m+1)!!
P_{m+1}^m(\xi_o){z\over R_p}\left({\varpi\over R_e}\right)^m.
\label{6.6}
\eeqa

\noindent Thus, the hydrodynamic potential $\delta U$ is given
by

\beq
\delta U=
\beta {z\over R_p}\left({\varpi\over R_e}\right)^m
e^{im\varphi}e^{i(\kappa-m)\Omega t}.
\label{6.7}
\eeq

\noindent Aside from the overall normalization then, the spatial
dependence of $\delta U$ is completely independent of the angular
velocity of the star!  A similar expression can be
obtained for the gravitational potential $\delta \Phi$ within the star. 
Using the fact that

\beqa
P_{m+1}^m&&(i\zeta)P_{m+1}^m(\mu)\nonumber \\&&=
(-1)^m(2m+1)!!
P_{m+1}^m(i\zeta_o){z\over R_p}\left({\varpi\over R_e}\right)^m,
\label{6.8}
\eeqa

\noindent the gravitational potential is given by

\beqa
\delta\Phi&& =
\alpha{z\over R_p}\left({\varpi\over R_e}\right)^m
e^{im\varphi}e^{i(\kappa-m)\Omega t}.
\label{6.9}
\eeqa

\noindent Thus the spatial dependencies of the potentials $\delta U$ and
$\delta\Phi$ are identical for the classical $r-$modes! This spatial
dependence can also be expressed in spherical coordinates as
$r^{m+1}Y_{m+1\,m}(\theta,\varphi)$, up to an overall normalization.  It
is interesting that this same function satisfies both the hydrodynamic
Eq.~(\ref{1.6}) and the gravitational potential Eq.~(\ref{1.7}). 

The Eulerian perturbation in the pressure for these modes is
determined from these two potentials by Eq.~(\ref{1.1}):

\beqa
{\delta p\over \rho} =
(\alpha+\beta)
{z\over R_p}\left({\varpi\over R_e}\right)^m
e^{im\varphi}e^{i(\kappa-m)\Omega t}.
\label{6.10}
\eeqa

\noindent We note that the constants $\alpha$ and $\beta$ used here have
been scaled by the factor $(-1)^m(2m+1)!!$ compared to their original
definitions in Eqs.~(\ref{3.6}) and (\ref{4.4}).  It is
also instructive to evaluate the Lagrangian perturbation of the
pressure, $\Delta p$, as defined in Eq.~(\ref{1.9}).  Using the
expressions in Eqs.~(\ref{6.7}) and (\ref{6.9}), we find that

\beqa
{\Delta p\over \rho}&& =
\Biggl\{ 1+{\alpha\over\beta} \nonumber\\
&& - {2(1-\zeta_o\cot^{-1}\zeta_o)[(2-\kappa)
(1+\zeta_o^2)-m\kappa\zeta_o^2]\over \kappa^2(2-\kappa)
\zeta_o[(1+3\zeta_o^2)\cot^{-1}\zeta_o-3\zeta_o]}\Biggr\}
\delta U.\nonumber\\
\label{6.11}
\eeqa

\noindent The term enclosed in $\{\}$ brackets in Eq.~(\ref{6.11})
depends only on the frequency of the mode $\kappa$, the angular
velocity of the star (through $\zeta_o$), and the amplitudes of the
perturbations $\alpha$ and $\beta$.  When the boundary condition
Eq.~(\ref{1.10}) (or equivalently \ref{4.13}) is satisfied, this term
vanishes.  Thus, we find that the Lagrangian perturbation in the
pressure $\Delta p$ vanishes identically for the classical $r-$modes
of the Maclaurin spheroids.  And this result (which is a consequence of
the extremely simple eigenfunctions for these modes) holds for stars
with {\it any} angular velocity!

\section{Discussion}
\label{sectionVII}

Our analysis, which follows closely in the footsteps of the remarkable
analysis of Bryan, provides several interesting insights into the
properties of the $r-$modes of rapidly rotating stars.  It demonstrates
for example, that these modes actually do exist in rapidly rotating
barotropic stars, and are not just figments of the first-order
perturbation theory (as claimed by some authors \cite{provost}).  This
analysis shows that some properties of the $r-$modes are not well
approximated by the low angular velocity expansions.  Figure~\ref{fig2}
illustrates, for example, that the first-order expression for the
angular velocity dependence of the frequency of the classical $r-$modes
is in fact superior to the second-order expression in the most rapidly
rotating stars.  This analysis shows that the frequency evolution of the
gravitational radiation emitted by the $r-$mode instability is likely to
be more interesting than had previously been thought~\cite{owen}. 
Figure~\ref{fig2} shows that the frequency of these modes will first
increase and then decrease as the angular velocity of the star is
reduced by the emission of gravitational radiation.  This analysis shows
that there is a much larger family of $r-$modes than the `classical'
$r-$modes studied for example by Papaloizou and Pringle~\cite{p&p}.  For
each pair of integers $l$ and $m$ (which satisfy $l\geq m\geq 0$) there
exist $l-m$ (or $l-1$ in the $m=0$ case) distinct $r-$modes.  This
analysis shows that a significant fraction of these previously unstudied
$r-$modes are subject to the gravitational radiation driven secular
instability.  This analysis has derived simple analytical expressions
for the eigenfunctions of the classical $r-$modes.  Both of the
potentials $\delta U$ and $\delta \Phi$ are proportional to
$r^{m+1}Y_{m+1\,m}(\theta, \varphi)$ for the classical $r-$modes in
Maclaurin spheroids of any angular velocity.  This analysis shows that
the Lagrangian variation in the pressure, $\Delta p$, associated with
the classical $r-$modes vanishes identically in Maclaurin spheroids of
arbitrary angular velocity.  Thus, the $r-$modes of the Maclaurin
spheroids provide a completely unsuitable model for the study of the
effects of bulk viscosity on the $r-$modes.  This analysis also provides
an interesting mathematical example of a hyperbolic eigenvalue problem. 
The $r-$modes studied here have the property that $\kappa^2<4$.  Thus,
the equation satisfied by the potential $\delta U$, Eq.~(\ref{1.6}), is
in fact hyperbolic.  Nevertheless, the boundary condition imposed on the
potential $\delta U$, Eq.~(\ref{1.10}), is of the mixed Dirchlet-Neumann
type that is generally associated with elliptic problems.  The
analytical solutions given here illustrate that this unusual hyperbolic
eigenvalue problem does nevertheless admit well behaved solutions. 


\acknowledgments

We particularly thank J.~Friedman, K.~Lockitch and N.~Stergioulas for
reading and criticizing the first draft of this manuscript.  We also
thank P.~Brady, C.~Cutler, B.~J.~Owen, and K.~S.~Thorne for helpful
discussions concerning this work; and we thank S.~Morsink for
providing us with her notes on the non-existence of pure axial modes
in the Maclaurin spheroids.  This research was supported by NSF grants
PHY-9408910 and PHY-9796079, and by NASA grants NAG5-3936 and
NAG5-4093.

\appendix
\section*{Bi-Spheroidal Coordinates}
\label{appendix}  

The coordinates $\xi$ and $\tilde{\mu}$ defined in Eqs.~(\ref{3.8})
through (\ref{3.10}) are rather unusual.  The purpose of this appendix
is to explore the geometrical properties of these coordinates.  The
coordinates $\xi$ and $\tilde{\mu}$ cover the planes usually described
with the spherical coordinates $r$ and ${\theta}$ or equivalently the
cylindrical coordinates $\varpi=\sqrt{x^2+y^2}$ and $z$.  First, we note
that it follows from Eqs.~(\ref{3.8}) through (\ref{3.10}) that

\beq
{\varpi^2\over 1-\xi^2}+{\kappa^2\over 4-\kappa^2}{z^2\over \xi^2}
= b^2.\label{a1}
\eeq

\noindent Thus, the surfaces of constant $\xi$ (with $\xi^2< 1$) are
spheroids for the $r-$modes which have $\kappa^2<4$.  The particular surface
$\xi=\xi_o$, with $\xi_o$ defined by

\beq
\xi_o^2= {a^2\zeta_o^2\over b^2}{\kappa^2\over 4-\kappa^2}
={\zeta_o^2\kappa^2\over 4(1+\zeta_o^2)-\kappa^2}<{\kappa^2\over 4}
<1,\label{a2}
\eeq

\noindent is identical to the surface of the star $\zeta=\zeta_o$.
Paradoxically, the constant $\tilde{\mu}$ surfaces also
satisfy the equation

\beq
{\varpi^2\over 1-\tilde{\mu}^2}+{\kappa^2\over 4-\kappa^2}{z^2\over 
\tilde{\mu}^2}
= b^2.\label{a3}
\eeq

\noindent Thus, the constant $\tilde{\mu}$ surfaces are the {\it same}
family of spheroids as the constant $\xi$ surfaces.  Thus the
coordinates $\xi$ and $\tilde{\mu}$ constitute a bi-spheroidal
coordinate system.  We note that the equatorial and polar radii of the
spheroid, $R_e$ and $R_p$ defined in Eqs.~(\ref{2.3.1}) and
(\ref{2.3.2}), are related to the constants $b$ and $\xi_o$ by

\beq
R_e^2 = b^2 (1-\xi_o^2),\label{a3.1}
\eeq

\beq
R_p^2 = b^2 \xi_o^2 {4-\kappa^2\over \kappa^2},\label{a3.2}
\eeq

In order to understand how the coordinates $\xi$ and $\tilde{\mu}$ 
cover the interior of the star it is helpful to introduce two additional
coordinates $s$ and $\tilde{\theta}$:

\beq
s^2\sin^2\tilde{\theta} = {\varpi^2\over R_e^2}
={(1-\xi^2)(1-\tilde{\mu}^2)\over 1-\xi_o^2},\label{a4}
\eeq

\beq
s^2\cos^2\tilde{\theta} = {z^2\over R_p^2} 
= {\xi^2\tilde{\mu}^2\over \xi_o^2}.\label{a5}
\eeq

\noindent The coordinate $s$ has the value 1 on the surface of the star
and 0 at its center.  The coordinate $\tilde{\theta}$ ranges from the
value 0 on the positive rotation axis, through $\pi/2$ on the equatorial
plane, to $\pi$ on the negative rotation axis.  Thus, the coordinates
$s$ and $\tilde{\theta}$ map the interior of the star into the unit
sphere in a natural way.  It will be instructive to express the
coordinates $\xi$ and $\tilde{\mu}$ then in terms of $s$ and
$\tilde{\theta}$.  These expressions can be obtained from
Eqs.~(\ref{a4}) and (\ref{a5}):

\beq
\xi^2 = \case{1}{2}(u + v),\label{a6}
\eeq

\beq
\tilde{\mu}^2 = \case{1}{2}(u-v),\label{a7}
\eeq

\noindent where

\beq
u = 1-s^2+s^2(\xi_o^2 + \cos^2\tilde{\theta}),\label{a8}
\eeq

\beq
v^2 = u^2 - 4s^2\xi_o^2\cos^2\tilde{\theta}.\label{a9}
\eeq

\noindent We note that while Eqs.~(\ref{a4}) and (\ref{a5}) are
symmetric in $\xi$ and $\tilde{\mu}$, this symmetry has been broken in
order to obtain the expressions (\ref{a6}) and (\ref{a7}). 

We will now show that the coordinates $\xi$ and $\tilde{\mu}$ also cover
the interior of the star when their values are restricted to the ranges:
$\xi_o\leq\xi\leq 1$ and $-\xi_o\leq\tilde{\mu}\leq \xi_o$.  It is easy
to see from the second equalities in Eqs.~(\ref{a4}) and (\ref{a5})
that each point $(\xi,\tilde{\mu})$ in the domain $\xi_o\leq\xi\leq
1$ and $-\xi_o\leq\tilde{\mu}\leq\xi_o$ corresponds to a point within
the star (i.e.  a point with $s\leq 1$).  Proving the converse is more
difficult.  We do this in three steps: First we show that the functions
$\xi$ and $\tilde{\mu}$ are real and finite at each point within the
interior of the star.  Second we show that these functions have no
critical points (e.g.  no maxima or minima) except on the surface of the
star.  Third, and last, we show that $\xi$ and $\tilde{\mu}$ are
confined to the ranges $\xi_o\leq\xi\leq 1$ and
$-\xi_o\leq\tilde{\mu}\leq \xi_o$ for points on the boundary. 

First we show that $\xi$ and $\tilde{\mu}$ are real and finite for each
point within the star.  The quantity $u$ defined in Eq.~(\ref{a8}) is
positive for points within the star (i.e.  points with $s\leq 1$).  Next we
show that $v$ (defined as the positive root in Eq.~\ref{a9}) is real and
thus positive for points within the star.  We do this by re-writing
$v^2$ as

\beqa
v^2=&&(1+s\xi_o\cos\tilde{\theta}+s\sqrt{1-\xi_o^2}\sin\tilde{\theta})\cr
&&\times(1+s\xi_o\cos\tilde{\theta}-s\sqrt{1-\xi_o^2}\sin\tilde{\theta})\cr
&&\times(1-s\xi_o\cos\tilde{\theta}+s\sqrt{1-\xi_o^2}\sin\tilde{\theta})\cr
&&\times(1+s\xi_o\cos\tilde{\theta}-s\sqrt{1-\xi_o^2}\sin\tilde{\theta}).
\label{a10}
\eeqa

\noindent Each of the terms on the right side of Eq.~(\ref{a10}) is
positive, since each is 1 plus the inner product of a pair of unit
vectors multiplied by $s$.  Thus $v^2\geq 0$ and so $v$ is real and
positive.  Further, $v\leq u$ and so $\xi^2$ and $\tilde{\mu}^2$ are
positive.  Thus $\xi$ and $\tilde{\mu}$ are finite and real for each
point of the interior of the star.  

Second we wish to show that the transformation between the bi-spheroidal
coordinates ($\xi,\tilde{\mu})$ and the standard cylindrical coordinates
$(\varpi,z)$ is non-singular everywhere within the star.  We do this by
evaluating the Jacobian matrix (i.e.  the matrix of partial derivatives)
of the transformation:

\beq
{\partial\xi\over\partial z} = {\kappa\tilde{\mu}(1-\xi^2)\over
b(\tilde{\mu}^2-\xi^2)\sqrt{4-\kappa^2}},\label{a11}
\eeq

\beq
{\partial\xi\over\partial\varpi}=
{\xi\sqrt{(1-\xi^2)(1-\tilde{\mu}^2)}\over b(\tilde{\mu}^2-\xi^2)},
\label{a12}
\eeq

\beq
{\partial\tilde{\mu}\over\partial z} =
-{\kappa\xi(1-\tilde{\mu}^2)\over b(\tilde{\mu}^2-\xi^2)\sqrt{4-\kappa^2}},
\label{a13}
\eeq

\beq
{\partial\tilde{\mu}\over\partial\varpi}=
-{\tilde{\mu}\sqrt{(1-\xi^2)(1-\tilde{\mu}^2)}\over
b(\tilde{\mu}^2-\xi^2)}.
\label{a14}
\eeq

\noindent These expressions show that the Jacobian matrix, and hence the
coordinate transformation between $(\xi,\tilde{\mu})$ and $(\varpi,z)$,
is non-singular except for the points within the star where $\xi=1$, or
$\xi^2=\tilde{\mu}^2=\xi_o^2$.  The transformation is also singular at
$\tilde{\mu}^2=1$, however, these points are not in the range of
interest to us here.  The non-singularity of the Jacobian matrix proves
that $\nabla_a\xi$ and $\nabla_a\tilde{\mu}$ are non-vanishing
everywhere within the star.  Thus, the maximum and minimum values of
$\xi$ and $\tilde{\mu}$ will only occur on the surface or the rotation
axis of the star. 

Third, and finally, we explore the values of the coordinates
$(\xi,\tilde{\mu})$ at specific physical locations in the star, e.g. 
the surface of the star, the rotation axis, etc.  We begin first with
the rotation axis.  In the $(s,\tilde{\theta})$ coordinates defined in
Eqs.~(\ref{a4}) and (\ref{a5}), the rotation axis corresponds to the
points where $\sin\tilde{\theta}=0$.  If follows then that these points
correspond to $\xi=1$ and $\tilde{\mu}^2=s^2\xi_o^2$.  Thus, the
rotation axis is the surface $\xi=1$, a singular surface of the
coordinate transformation.  The coordinate $\tilde{\mu}$ takes on its
entire range, $-\xi_o\leq \tilde{\mu}\leq\xi_o$, for points along this
axis.  The equatorial plane of the star, $\cos\tilde{\theta}=0$,
corresponds to the coordinate surface $\tilde{\mu}=0$.  The coordinate
$\xi^2=1-s^2+s^2\xi_o^2$ ranges from $\xi=1$ (on the rotation axis) to
the value $\xi=\xi_o$ on the surface of the star. 

The surface of the star, $s=1$, is unexpectedly complicated in the
$(\xi,\tilde{\mu})$ coordinate system.  For points of the stellar
surface near the equator, $\cos^2\tilde{\theta}\leq \xi_o^2$, the
function $v$ defined in Eq.~(\ref{a9}) has the value:
$v=\xi_o^2-\cos^2\tilde{\theta}$.  Thus, the surface of the star in this
region has $\xi=\xi_o$ while $\tilde{\mu}=\cos\tilde{\theta}$.  In the
regions of the stellar surface near the rotation axis,
$\cos^2\tilde{\theta}\geq\xi_o^2$, the function $v$ (defined as the {\it
positive} root in Eq.~\ref{a9}) has the value:
$v=\cos^2\tilde{\theta}-\xi_o^2$.  Thus the stellar surface in these
regions have $\tilde{\mu}=\pm\xi_o$ and $\xi^2=\cos^2\tilde{\theta}$. 
The role of $\xi$ and $\tilde{\mu}$ as ``radial'' and ``angular''
coordinates are reversed therefore in different regions of the star. 

In summary then, we have shown that the maximum and minimum values of
$\xi$ are $1$ and $\xi_o$ respectively, while the maximum and minimum
values of $\tilde{\mu}$ are $\pm\xi_o$.  This concludes our
demonstration that the coordinates $(\xi,\tilde{\mu})$ when confined to
the ranges $\xi_o\leq \xi \leq 1$ and $-\xi_o\leq\tilde{\mu}\leq\xi_o$
do form a non-singular coordinate system that covers the interior of the
star.  The transformation between these and the usual cylindrical
coordinates is singular only on the rotation axis, $\xi=1$, and at the
singular point $\xi^2=\tilde{\mu}^2=\xi_o^2$ on the surface of the star. 

Finally, it will be useful to work out the relationship between the
surface values of the $(\xi,\tilde{\mu})$ coordinates with those of the
oblate spheroidal coordinates $(\zeta,\mu)$.  Using the definitions of
the $(s,\tilde{\theta})$ coordinates introduced in Eqs.~(\ref{a4}) and
(\ref{a5}), and the definitions of the oblate spheroidal coordinates
$(\zeta,\mu)$ from Eqs.~(\ref{3.1}), (\ref{3.2}), and (\ref{3.3}), it is
straightforward to show that

\beq
s^2\sin^2\tilde{\theta}={(\zeta^2+1)(1-\mu^2)\over \zeta_o^2+1},\label{a15}
\eeq

\beq
s^2\cos^2\tilde{\theta}={\zeta^2\mu^2\over\zeta_o^2}.\label{a16}
\eeq

\noindent Thus on the surface of the star, $\zeta=\zeta_o$ and $s=1$, we
have $\cos\tilde{\theta}=\mu$.  We have also shown that
$\tilde{\mu}=\cos\tilde{\theta}$ on the portion of the surface of the
star where $\cos^2\tilde{\theta}\leq \xi_o^2$, and that
$\xi=\pm\cos\tilde{\theta}$ on the portion of the surface where
$\cos^2\tilde{\theta}\geq\xi_o^2$.  Thus we find that $\tilde{\mu}=\mu$
on the portion of the surface where $\xi=\xi_o$, and that $\xi=\pm\mu$
on the portion of the surface where $\tilde{\mu}=\pm\xi_o$.



\begin{references}
\bibitem{g-modes} The Maclaurin spheroids are a special case of
barotropic stellar models: stars in which the density of the perturbed
fluid depends only on the pressure.  In non-barotropic stars there
also exist buoyancy forces that dominate the behavior of a class of
modes called $g-$modes.  The non-barotropic analogues of the
generalized $r-$modes discussed here will almost certainly be
influenced at some level by buoyancy forces.  At sufficiently small
angular velocities (i.e. at angular velocities smaller than the
Brunt-V\"ais\"al\"a frequency) buoyancy forces could well dominate the
dynamics of some of these modes; and such modes might then be called
generalized $g-$modes in non-barotropic models.  By analogy some might
prefer to call these modes generalized $g-$modes even in the barotropic
case.  Since we do not at present know which (if any) of these modes
might be dominated by buoyancy forces in the non-barotropic case, we
prefer to refer to them here as $r-$modes.  In the barotropic case the
dynamics of these modes are dominated by rotational forces.
\bibitem{l=m+1} It has been traditional in the literature to discuss
the $r-$modes in terms of a vector-spherical-harmonic decomposition of
the Lagrangian displacement.  In terms of that traditional
decomposition, the `classical' $r-$modes correspond to the case $l=m$.
The analysis presented here uses a certain scalar potential $\delta U$
from which the Lagrangian displacement is determined.  The analytic
solutions that we find for $\delta U$ are particular spheroidal
harmonic functions.  The classical $r-$modes correspond to the case
$l=m+1$ in terms of these spheroidal harmonics.  The potential $\delta
U$ for the classical $r-$modes can also be expressed as a spherical
harmonic with $l=m+1$.  In general, however, the modes discussed here
do not have simple finite representations in terms of spherical
harmonics.
\bibitem{note2} The Legendre polynomial $P_l$ has $l$ distinct roots,
all of which lie in the interval $(-1,1)$~\cite{bateman}.  It follows
that $dP_l/d\xi$ has $l-1$ roots which interleave the roots of $P_l$,
while $d^{\,m}P_l/d\xi^m$ has $l-m$ roots which interleave those of
$d^{\,m-1}P_l/d\xi^{m-1}$, etc.  For the case $m=0$ then, the roots of
Eq.~(\ref{5.111}) are simply the roots of $dP_l/d\xi$.  Thus, there are
$l-1$ distinct roots for $\kappa_o$ in the $m=0$ case, and these lie in
the interval $-2<\kappa_o<2$.  For the case $m>0$ the first term in
Eq.~(\ref{5.111}) has $l-m$ distinct zeros, while the second term has
$l-m$ (including the one at $\kappa_o=2$) zeros that interleave those of
the first.  Evaluating the right side of Eq.~(\ref{5.111}) at these
zeros of its second term, we find that its sign alternates.  Thus,
the right side of Eq.~(\ref{5.111}) has $l-m$ zeros that interleave
those of its second term.  Thus, these zeros lie in the range
$-2<\kappa_o<2$.  These are all of the zeros of the right side of
Eq.~(\ref{5.111}) because it is a polynomial of degree $l-m$.
\end{references}
\end{document}